\documentstyle[prl,aps,psfig,floats]{revtex}

\begin{document}

\wideabs{
\title{Exact singlet bond ground states for electronic models}
\author{D.V.Dmitriev, V.Ya.Krivnov and A.A.Ovchinnikov}
\address{Max-Planck-Institut fur Physik Komplexer Systeme, 
Nothnitzer Str. 38, 01187 Dresden, Germany \\
and Joint Institute of Chemical Physics of RAS,
Kosygin str.4, 117977, Moscow, Russia.}
\maketitle
\begin{abstract}
We have proposed several 1D and 2D electronic models with the exact ground
state. The ground state wave function of these models is represented in terms 
of "singlet bond" functions consisting of homopolar and ionic configurations.
The Hamiltonians of the models include the correlated hopping of electrons,
pair hopping terms and the spin interactions. One of the 1D models 
demonstrates spontaneous breaking of the translational symmetry. We studied
also the model describing the transition point between the phases with and
without an off-diagonal long-range order.
\end{abstract}


} 


The study of strongly correlated electron systems has been an important
subject in theoretical condensed matter physics. In general, the Hamiltonians
of these systems include many types of interactions, and they are difficult to 
solve. The integrable models provided us a very good understanding of the
correlation effects in many-body systems. Unfortunately, the construction of
such models is difficult due to the strict conditions for the integrability.
In recent years there has been increasing interest in studying models 
where at least the ground state can be found exactly \cite{strack,boer,Z,mik}.
The most popular methods for the construction of exact ground state are
so-called optimal ground state (OGS) approach \cite{boer} and the matrix-product 
(MP) method \cite{Z,mik}. The ground state wave function in the MP method is
represented by $Trace$ of a product of matrices describing single-site states. 
This ground state is optimal in the sense that it is the ground state
of each local interaction. This method allows to construct a large class of
spin models. The similar spirit has been used in the OGS method for construction
of the electronic models with the special ground states.

In this Letter we propose new 1D and 2D models of interacting electrons with
the exact ground state. We note that our models have ground states which are
very different from those constructed in the OGS approach.
The ground state wave function of our models is
expressed in terms of the two-particle {\it "singlet bond"} (SB) function
located on sites $i$ and $j$ of the lattice:
\begin{equation}
[i,j] = c_{i,\uparrow}^{+} c_{j,\downarrow}^{+}
- c_{i,\downarrow}^{+} c_{j,\uparrow}^{+}
+ x\,(c_{i,\uparrow}^{+} c_{i,\downarrow}^{+}
+ c_{j,\uparrow}^{+} c_{j,\downarrow}^{+}) \left| 0 \right \rangle ,
\label{SB}
\end{equation}
where $c_{i,\sigma}^{+}$, $c_{i,\sigma}$ are the Fermi operators and $x$ is an
arbitrary coefficient. The SB function is the generalization of the RVB
function \cite{PWA} including ionic states. The presence of the ionic states
is very important from the physical point of view because, as a rule, the bond 
functions contain definite amount of the ionic states as well.

It is known a set of 1D and 2D quantum spin models the exact ground state of
which can be represented in the RVB form \cite{MG,SS,AKLT,Japan}. It is
natural to try to find electronic models with exact ground state at
half-filling formed by SB functions in the same manner as for above mentioned
spin models. The electronic models of these types include the correlated
hopping of electrons as well as the spin interactions and pair hopping terms.


{\it The model with dimerization}.--
As the first example we consider the 1D electronic model with the two-fold
degenerate ground state in the form of the simple product of SB dimers, 
similarly to the ground state of the well-known spin-$\frac 12$ Majumdar-Ghosh
model \cite{MG}. For the half-filling case the proposed ground state wave
functions are:
\begin{equation}
\Psi_1 = [1,2][3,4]...[N-1,N]
\label{mgwf1} 
\end{equation}
and
\begin{equation}
\Psi_2 = [2,3][4,5]...[N,1]
\label{mgwf2}
\end{equation}

In order to find the Hamiltonian for which the wave functions (\ref{mgwf1})
and (\ref{mgwf2}) are the exact ground state wave functions, we represent the 
Hamiltonian as a sum of local Hamiltonians $h_i$ defined on three neighboring
sites (periodic boundary conditions are supposed): 
\begin{equation}
H = \sum_{i=1}^N h_i
\label{h}
\end{equation}

The basis of three-site local Hamiltonians $h_i$ consists of 64 states, while
only eight of them are present in $\Psi_1$ and $\Psi_2$. These 8 states are
\begin{equation}
[i,i+1]\,\varphi _{i+2}, \qquad \varphi _{i}\,[i+1,i+2] ,
\label{mggs}
\end{equation}
where $\varphi _{i}$ is one of four possible electronic states of
$i$-th site: $\left| 0\right \rangle _i$, $\left|\uparrow\right\rangle _i$, 
$\left|\downarrow \right \rangle _i$, $\left| 2 \right \rangle _i$.

The local Hamiltonian $h_i$ for which all the functions (\ref{mggs})
are the exact ground state wave functions can be written as the sum of the
projectors onto other 56 states $\left|\chi_k \right \rangle$
\begin{equation}
h_i = \sum_{k} \lambda _{k}  
\left|\chi_k \right \rangle  \left \langle \chi_k \right| ,
\label{hn}
\end{equation}
where $\lambda_{k}$ are arbitrary positive coefficients.
This means that the 
wave functions $\Psi_1$ and $\Psi_2$ are the ground states of each local
Hamiltonian with zero energy. Hence, $\Psi_1$ and $\Psi_2$ are the ground
state wave functions of $H$ with zero energy.

In general case, the local Hamiltonian $h_i$ is many-parametrical and depends
on parameters $\lambda _k$ and $x$. We consider one of the simplest forms of
$h_i$ including the correlated hopping of electrons of different types and
spin interactions between nearest- and next-nearest neighbor sites:
\begin{eqnarray}
h_i &=& 2 - x\,(t_{i,i+1}+t_{i+1,i+2}) \nonumber \\
&+&  (x^2-(1+x^2)(1-n_{i+1})^2)\;T_{i,i+2} \nonumber \\
&+& 8\frac{1-x^2}{3}({\bf S}_{i}\cdot {\bf S}_{i+1}+{\bf S}_{i+1}\cdot {\bf
S}_{i+2}+{\bf S}_{i}\cdot {\bf S}_{i+2}) ,
\label{mgh} 
\end{eqnarray}
where
\begin{eqnarray}
T_{i,j} &=& \sum_{\sigma}(c_{i,\sigma}^{+}c_{j,\sigma}+c_{j,\sigma}^{+}c_{i,\sigma})
(1-n_{i,-\sigma}-n_{j,-\sigma}) , \nonumber \\
t_{i,j} &=& \sum_{\sigma}   (c_{i,\sigma}^{+}c_{j,\sigma}+
c_{j,\sigma}^{+}c_{i,\sigma})(n_{i,-\sigma}-n_{j,-\sigma})^2
\nonumber 
\end{eqnarray}
and ${\bf S}_i$ is the $SU(2)$ spin operator.

Each local Hamiltonian $h_i$ is a non-negatively defined operator at $|x|\le 1$.
The following statements related to the Hamiltonian (\ref{mgh}) are valid:

1. The functions (\ref{mgwf1}) and (\ref{mgwf2}) are the only two ground state
wave functions of the Hamiltonian (\ref{mgh}) at $N_e = N$ ($N_e$ is the total
number of electrons).  They are not orthogonal, but their overlap is $\sim
e^{-N}$ at $N\gg 1$. 

2. The ground state energy $E_0(N_e/N)$ is a symmetrical function with respect
to the point $N_e/N=1$ and has a global minimum $E_0=0$ at $N_e/N=1$.

3. The translational symmetry of (\ref{mgh}) is spontaneously broken in the
ground state leading to the dimerization:
\[
\left \langle |t_{i,i+1}-t_{i+1,i+2}| \right \rangle = 2
\]

The excited states of the model can not be calculated exactly but we expect
that there has to be a gap, because the ground state is formed by the
ultrashort-range SB functions. If it is the case, the function $E_0(N_e/N)$
has a cusp at $N_e/N = 1$.

Actually, this model is the fermion version of the  Majumdar -- Ghosh spin
model. Moreover, it reduces to the Majumdar -- Ghosh model at $x=0$ and in the 
subspace with $n_i=1$.

\begin{figure}[t]
\unitlength1cm
\begin{picture}(8,4)
\centerline{\psfig{file=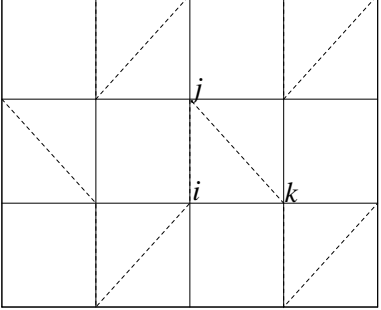,angle=-90,width=5cm}}
\end{picture}
\caption{ \label{sss} The lattice of the Shastry -- Sutherland model}
\end{figure}

For $x=1$ the Hamiltonian (\ref{mgh}) simplifies and takes the form:
\begin{equation}
H = -2 \sum_{j} (t_{j,j+1}-1) - \sum_{j} e^{i\pi n_{j+1}} T_{j,j+2}
\label{mgh1}
\end{equation}
%
%
%
{\it The 2D model}.--
We can easily construct the 2D electronic model with the exact ground state
which is analogous to the Shastry -- Sutherland model \cite{SS} (Fig.1). The
Hamiltonian of this model is:
\begin{equation}
H = \sum_{\{i,j,k\}} h_{i,j} + h_{i,k} + h^d_{j,k} ,
\label{ssh}
\end{equation}
where the sum is over all triangles $\{i,j,k\}$, one of which is shown on
Fig.1. So, each diagonal line belongs to the two different triangles.
The local Hamiltonians $h^d_{j,k}$ acting on the diagonal of the triangle
$\{i,j,k\}$, and $h_{i,j}$, $h_{i,k}$ have the form (for the sake of simplicity
we put $x=1$)
\begin{eqnarray}
h^d_{j,k} &=& -2\;t_{j,k} + 4 \nonumber \\
h_{i,j}   &=& - t_{i,j} - e^{i\pi n_k}\;T_{i,j}  \nonumber \\
h_{i,k}   &=& - t_{i,k} - e^{i\pi n_j}\;T_{i,k}  \nonumber
\end{eqnarray}

It is easy to check that
\[
h^d_{j,k}\; \left| \varphi _{i}\;[j,k] \right\rangle
= (h_{i,j}+h_{i,k}) \;\left| \varphi _{i}\;[j,k]\right\rangle = 0   
\]

All other states of the local  Hamiltonian $h_{i,j}+h_{i,k}+h^d_{j,k}$ have
higher energies. Therefore, the ground state wave function in the half-filling
case is the product of the SB functions located on the diagonals shown by
dashed lines on Fig.1. This model has non-degenerate singlet ground state with
ultrashort-range correlations.


{\it The ladder model}.--
Let us now consider electronic models with a more complicated ground state
including different configurations of short-range SB functions. The form
of these ground states is similar to that for spin models proposed in
\cite{AKLT} and generalized in \cite{JETP}. In the 1D case our model describes
two-leg ladder model (Fig.2). Its ground state is a superposition of the SB
functions where each pair of nearest neighbor rungs of the ladder is connected
by one SB. One of possible configurations of singlet bonds is shown on Fig.2. 

\begin{figure}[t]
\unitlength1cm
\begin{picture}(11,2)
\centerline{\psfig{file=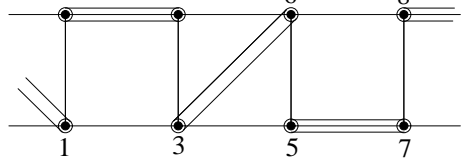,angle=-90,width=6cm}}
\end{picture}
\caption{ \label{sbbs} The two-leg ladder model.}
\end{figure}

The wave function of this ground state can be written as:
\begin{equation} 
\Psi _{s} = \psi ^{\lambda \mu }(1)g_{\mu \nu }\psi ^{\nu \rho }(2)g_{\rho
\kappa }\ldots \psi ^{\sigma \tau}(N)g_{\tau \lambda }
\label{sbs}
\end{equation}

The wave function of this type for spin models has been proposed in \cite{AKLT}.
The functions $\psi ^{\lambda \mu }(i)$ describes $i$-th rung of the ladder
\begin{equation} 
\psi ^{\lambda \mu }(i) = c_1 \varphi _{2i-1}^{\lambda} \varphi _{2i}^{\mu } 
+ c_2 \varphi _{2i}^{\lambda} \varphi _{2i-1}^{\mu } 
\label{rung}
\end{equation}
with
\[
\varphi _{k}^{\lambda} = \left (
\begin{array}{c}
\left|\uparrow   \right \rangle _k \\
\left|\downarrow \right \rangle _k \\
\left| 2   \right \rangle _k\\
\left| 0   \right \rangle _k
\end{array} \right)
,\qquad
g_{\lambda \mu } = \left (
\begin{array}{cccc}
 0 & 1 & 0 & 0 \\
-1 & 0 & 0 & 0 \\
 0 & 0 & 0 & x \\
 0 & 0 & x & 0
\end{array}\right)
\]

It is easy to see that
\[
g_{\lambda \mu }\varphi _{i}^{\lambda}\varphi _{j}^{\mu} = [i,j]
\]

Therefore, the function $\Psi_s$ is a singlet wave function depending on two
parameters $x$ and $c_1/c_2$. Actually, this form of $\Psi_s$ is equivalent to
the MP form with $4\times 4$ matrices $A_{\lambda \nu }(i) =
g_{\lambda\mu}\psi ^{\mu\nu}(i)$. Moreover, at $x=0$ and $c_1/c_2=-1$ the
function $\Psi_s$ reduces to the wave function of the well-known AKLT spin-$1$
model.

In order to find the Hamiltonian for which the wave function (\ref{sbs}) is
the exact ground state wave function, it is necessary to consider what states
are present on the two nearest rungs in the $\Psi _s$. It turns out that there
are only 16 states from the total 256 ones in the product $\psi ^{\lambda \mu
}(i)g_{\mu \nu }\psi ^{\nu \rho }(i+1)$. The local Hamiltonian $h_i$ acting on
two nearest rungs $i$ and $i+1$ can be written in the form of (\ref{hn}) with
the projectors onto the 240 missing states. The total Hamiltonian is the sum
of local ones (\ref{h}). The explicit form of this Hamiltonian is very
cumbersome and, therefore, it is not given here.

The correlation functions in the ground state (\ref{sbs}) can be calculated
exactly in the same manner as it was done for spin models \cite{AKLT,pr}. It
can be shown that all of correlations exponentially decay in the ground state.
We expect also that this model has a gap.

This method of construction of the exact ground state can be generalized also
to 2D and 3D lattices \cite{JETP}. Following \cite{JETP}, one can rigorously
prove that the ground state of these models is always a non-degenerate singlet.


{\it 1D models with the giant spiral order}.--
There is one more spin-$\frac 12$ model with an exact ground state of the RVB
type \cite{Japan}. Its Hamiltonian has the form
\begin{equation}
H=-\sum_{i}{\bf S}_{i}\cdot {\bf S}_{i+1} + \frac{1}{4}
\sum_{i}{\bf S}_{i}\cdot {\bf S}_{i+2}  \label{spinh}
\end{equation}
This model describes the ferromagnet--antiferromagnet transition point \cite{pr}.
The exact singlet ground state can be expressed by the combinations of the RVB
functions $(i,j)$ distributed uniformly over the  lattice points:
\begin{equation}
\Phi _0 =\sum (i,j)(k,l)(m,n)\ldots \;\;,
\label{swf}
\end{equation}
where the summation is done over all combinations of sites under the condition 
that $i<j,\;k<l,\;m<n\ldots$. The spin correlations in the singlet ground
state show giant spiral structure \cite{Japan,pr}.


The analog of the wave function (\ref{swf}) in the SB terms is:
\begin{equation}
\Psi _0 =\sum_{i<j\ldots} (-1)^P\;[i,j][k,l][m,n]\ldots  ,
\label{e1wf}
\end{equation}
where $P=(i,j,k,l,\ldots)$ is the permutation of numbers $(1,2,\ldots N)$.
It is interesting to note that the singlet wave function (\ref{e1wf}) can be
also written in the MP form but with an infinite size matrices \cite{prb}.

In order to find the Hamiltonian for which the wave function (\ref{e1wf}) 
is the exact ground state wave function, let us consider what 
states are present on the two nearest sites in the $\Psi _0$.
It turns out \cite{prb} that there are only 9 states from the total 16 ones
in (\ref{e1wf}). They are
\begin{eqnarray}
&&\left|\uparrow \uparrow \right \rangle ,\qquad
\left|\downarrow \downarrow  \right \rangle ,\qquad
\left|\uparrow \downarrow + \downarrow \uparrow  \right \rangle ,
\qquad \left|20-02 \right \rangle , 
\nonumber \\
&&\left|\uparrow \downarrow - \downarrow \uparrow  \right \rangle
- x \: \left|20+02 \right \rangle ,\qquad
\left|\uparrow 0 - 0\uparrow \right \rangle ,\nonumber \\
&&\left|\uparrow 2 - 2\uparrow \right \rangle ,\qquad
\left|\downarrow 0 - 0\uparrow \right \rangle ,\qquad
\left|\downarrow 2 - 2\downarrow \right \rangle 
\label{states1}
\end{eqnarray}

The local Hamiltonian $h_{i,i+1}$ can be written as the sum of the projectors 
onto the 7 missing states (\ref{hn}). 
At $|x|>1$ the most simple form of this total Hamiltonian is:
\begin{eqnarray}
H &=& \sum_{i=1}^{N} \left(
T_{i,i+1} + \frac{2}{x} t_{i,i+1}
-4\:{\bf S}_{i}\cdot {\bf S}_{i+1} \right.
\nonumber \\
&+& \left. \frac{4}{x^2}{\bf \eta}_{i}\cdot {\bf \eta}_{i+1}
+ 4\,\frac{x^2-3}{x^2}\eta_{i}^z\eta_{i+1}^z + 1 \right)
\label{e1h}
\end{eqnarray}
We use here ${\bf \eta}$ operators:
\[
\eta_i^{+}=c_{i,\downarrow }^{+}\:c_{i,\uparrow }^{+}, \qquad
\eta_i^{-}=c_{i,\uparrow }\:c_{i,\downarrow }, \qquad
\eta_i^{z}=\frac {1-n_i}{2} , 
\]
which form another $SU(2)$ algebra \cite{yang,eta}, and 
${\bf \eta}_{1}\cdot {\bf \eta}_{2}$ is a scalar product of pseudo-spins
${\bf \eta}_{1}$ and ${\bf \eta}_{2}$. We note that this Hamiltonian commutes
with ${\bf S}^2$, but does not commute with ${\bf \eta}^2$.
It can be proved \cite{prb} that only three multiplets are the ground states of
(\ref{e1h}): the singlet state (\ref{e1wf}), the trivial ferromagnetic state
$S=N/2$ and the state with $S=N/2-1$.

The norm and the correlators of the electronic model (\ref{e1h}) in the
singlet ground state are exactly calculated \cite{prb}. For example, the norm
of (\ref{e1wf}) is:
\[
\langle \Psi _0| \Psi _0\rangle  = \left. \frac {d^{N}}{d\xi ^{N}} 
\left( 2\frac{1+\cosh(x \xi )}{\cos^2(\frac{\xi}{2})}  \right)  \right|_{\xi=0}
\]

The correlators at $N\gg 1$ are
\begin{eqnarray}
\langle \eta_i^z \eta_{i+l}^z \rangle &=& O\left(\frac {1}{N^2}\right) 
,\qquad
\langle \eta_i^+ \eta_{i+l}^- \rangle = O\left(\frac {1}{N^2}\right) ,
\nonumber \\
\langle c_{i,\sigma}^{+}\:c_{i+l,\sigma}\rangle &=& O\left(\frac 1N\right) ,
\qquad
\left\langle {\bf S}_{i}{\bf S}_{i+l}\right\rangle =
\frac{1}{4}\cos \left( \frac{2\pi l}{N}\right)
\label{spincos1} 
\end{eqnarray}

The correlator $\langle \eta_i^{+} \eta_{i+l}^{-} \rangle $ which determines
the off-diagonal long-range order (ODLRO) \cite{yang} vanishes in the
thermodynamic limit. At the same time the spin-spin correlations have a spiral 
form, and the period of the spiral equals to the system size as in the spin 
model (\ref{spinh}).


Another electronic model can be obtained by making the canonical
transformation $c^{+}_{i,\uparrow}\to c^{+}_{i,\uparrow}$ and
$c^{+}_{i,\downarrow}\to c_{i,\downarrow}$. As a result of this
transformation, the SB function (\ref{SB}) becomes:
\begin{equation}
\{i,j\} = c_{i,\uparrow}^{+} c_{i,\downarrow}^{+}
- c_{j,\uparrow}^{+} c_{j,\downarrow}^{+}
+ x\,(c_{i,\uparrow}^{+} c_{j,\downarrow}^{+}
+ c_{i,\downarrow}^{+} c_{j,\uparrow}^{+} ) \left| 0 \right \rangle ,
\label{SB0}
\end{equation}
and the wave function (\ref{e1wf}) changes to
\begin{equation}
\Psi _0 =\sum_{i<j\ldots} \{i,j\}\{k,l\}\{m,n\}\ldots
\label{e2wf}
\end{equation}

The function (\ref{e2wf}) for $|x|>1$ is the exact ground state wave function
of the transformed Hamiltonian:
\begin{eqnarray}
H &=& \sum_{i=1}^{N} \left(
-T_{i,i+1} + \frac{2}{x} t_{i,i+1}^{\prime}
-4{\bf \eta}_{i}\cdot {\bf \eta}_{i+1} \right.
\nonumber \\
&+& \left.\frac{4}{x^2}{\bf S}_{i}\cdot {\bf S}_{i+1}
+4\,\frac{x^2-3}{x^2} S_{i}^z S_{i+1}^z +1 \right) ,
\label{e2h}
\end{eqnarray}
where
\[
t_{i,i+1}^{\prime} = \sum_{\sigma} \sigma(c_{i,\sigma}^{+}c_{i+1,\sigma}+
c_{i+1,\sigma}^{+}c_{i,\sigma})(n_{i,-\sigma}-n_{i+1,-\sigma})^2
\]
This Hamiltonian commutes with ${\bf \eta}^2$ but does not commute with ${\bf
S}^2$. Therefore, the eigenfunctions of the Hamiltonian (\ref{e2h}) can be
described by quantum numbers ${\bf\eta}$ and $\eta^z$. For the cyclic model 
(\ref{e2h}) the states with three different values of $\eta$ have zero 
energy \cite{prb} [as it was for the model (\ref{e1h})]. They include
one state with $\eta =0$ (\ref{e2wf}), all states with $\eta =N/2$:
\begin{equation}
\Psi_{N/2,\;\eta^z} = (\eta^+)^{N/2-\eta^z} \left|0 \right \rangle ,
\label{e2wf2}
\end{equation}
and the states with $\eta =N/2-1$.
Therefore, for the case of one electron per site ($\eta^z=0$), the ground 
state of the model (\ref{e2h}) is three-fold degenerate.

The correlation functions in the ground states with $\eta =N/2$ and $\eta
=N/2-1$ for the half-filling case coincide with each other and at $N\gg 1$
they are:
\begin{eqnarray}
\langle c_{i,\sigma}^{+}\:c_{i+l,\sigma}\rangle = O\left(\frac 1N\right) , 
\qquad
\langle {\bf S}_{i}{\bf S}_{i+l} \rangle = O\left(\frac {1}{N^2}\right) ,
\nonumber \\
\left\langle \eta_i^z \eta_{i+l}^z \right\rangle = O\left(\frac 1N\right) ,
\qquad
\left\langle \eta_i^- \eta_{i+l}^+ \right\rangle = \frac{1}{4} + 
O\left(\frac 1N\right) 
\label{etaferro} 
\end{eqnarray}
The existence of the ODLRO immediately follows from the latter equations.
The correlation functions in the ground state (\ref{e2wf}) have similar 
forms as in Eqs.(\ref{spincos1}):
\begin{eqnarray}
&&\langle c_{i,\sigma}^{+}\:c_{i+l,\sigma}\rangle = O\left(\frac {1}{N}\right),
\qquad 
\langle {\bf S}_{i}{\bf S}_{i+l} \rangle = O\left(\frac {1}{N^2}\right),
\nonumber \\
&&\left\langle \eta_i^- \eta_{i+l}^+ \right\rangle =
2 \left\langle \eta_i^z \eta_{i+l}^z \right\rangle =
\frac{1}{6}\cos \left( \frac{2\pi l}{N}\right) 
\label{etacos} 
\end{eqnarray}
The giant spiral ordering in the last equation implies the existence of the
ODLRO and, therefore, the superconductivity \cite{yang} in the ground state 
(\ref{e2wf}). 

Similarly to the original spin model (\ref{spinh}) \cite{pr,KO} the last two
electronic models (\ref{e1h}),(\ref{e2h}) describe the transition points
on the phase diagram between the phases with and without a long-range order
[ferromagnetic for the model (\ref{e1h}) and off-diagonal for the model
(\ref{e2h})]. Therefore, we suggest the formation of the ground state with
giant spiral order (ferromagnetic or off-diagonal) as a probable scenario of
the subsequent destruction of the ferromagnetism and superconductivity.


In summary, we have constructed electronic models with an exact ground state.
The ground state wave function of these models is built from SB functions in
the same manner as the well-known RVB ground states of spin models. We have
considered three types of SB ground states. However, we note that the proposed
approach can be generalized for the construction of other models with the
ground states of more complicated SB forms.


Authors are grateful to Max-Planck-Institut fur Physik Komplexer Systeme for a 
kind hospitality. This work was supported by RFFR.

\end{document}